\documentclass[aps,preprint,floatfix]{revtex4-1}
\usepackage{graphicx}
\usepackage{amsmath,amssymb,bm}
\begin{document}
%%%%%%%%%%%%%%%%%%%%%%%%%%%%%%%%%%%%%%%%%%

\title{Collapse of superhydrophobicity on nanopillared surfaces}
%%%%%%%%%%%%%%%%%%%%%%%%%%%%%%%%%%%%%%%%%%

\author{Matteo Amabili}
\affiliation{Sapienza Universit\`a di Roma, Dipartimento di Ingegneria
Meccanica e Aerospaziale, 00184 Rome, Italy}
\author{Alberto Giacomello}
\email[]{alberto.giacomello@uniroma1.it}
\affiliation{Sapienza Universit\`a di Roma, Dipartimento di Ingegneria
Meccanica e Aerospaziale, 00184 Rome, Italy}
\author{Simone Meloni}
\affiliation{Sapienza Universit\`a di Roma, Dipartimento di Ingegneria
Meccanica e Aerospaziale, 00184 Rome, Italy}
\author{Carlo Massimo Casciola}
\affiliation{Sapienza Universit\`a di Roma, Dipartimento di Ingegneria
Meccanica e Aerospaziale, 00184 Rome, Italy}

\date{\today}

%%%%%%%%%%%%%%%%%%%%%%%%%%%%%%%%%%%%%%%%%%%%%%%%%%%%%%%%%%%%%

\begin{abstract}
The mechanism of the collapse of the superhydrophobic state is
elucidated for submerged nanoscale textures forming a three-dimensional
interconnected vapor domain. This key issue for the design of nanotextures poses significant
simulation challenges as it is characterized by diverse time and
length scales.  State-of-the-art atomistic rare events simulations are
applied for overcoming the long timescales connected with the large free
energy barriers. In such interconnected surface cavities wetting starts
with the formation of a liquid finger between two
pillars. This break of symmetry induces a more gentle bend in the rest
of the liquid-vapor interface, which triggers the wetting of the
neighboring pillars. This collective mechanism, involving the wetting of several pillars at the same time, 
could not be captured by previous atomistic simulations using surface models comprising a small number of pillars (often just one). Atomistic
results are interpreted in terms of a sharp-interface continuum model
which suggests that line tension, condensation, and other nanoscale
phenomena play a minor role in conditions close to coexistence. 
\end{abstract}

%%%%%%%%%%%%%%%%%%%%%%%%%%%%%%%%%%%%%%%%%%%%%%%%%%%%%%%%%%%%%

\maketitle

%%%%%%%%%%%%%%%%%%%%%%%%%%%%%%%%%%%%%%%%%%%%%%%%%%%%%%%%%%%%%
%INTRO
\section{Introduction}
%%%%%%%%%%%%%%%%%%%%%%%%%%%%%%%%%%%%%%%%%%%%%%%%%%%%%%%%%%%%%

Superhydrophobicity stems from the presence of a gaseous layer between a body
of liquid and a surface. This suspended ``Cassie'' state is fostered by surface
roughness and hydrophobic coatings \cite{lafuma2003}; the composite
liquid-gas-solid interface results in important properties for submerged and
``dry'' technological applications, such as self-cleaning, enhanced liquid
repellency, and drag reduction \cite{Quere2008,rothstein2010}. The
superhydrophobic Cassie state is sustained by capillary forces, which can be
overcome by variations in the liquid pressure, temperature, or other external
forces which trigger the transition to the fully wet Wenzel state.  Thus, the
design of textured surfaces and coatings has the aim of realizing a robust
Cassie state.

The stability of the Cassie state depends not only on the thermodynamic
conditions but also on the surface geometry and chemistry, which can be
engineered in order to achieve robust superhydrophobicity.  Hydrophobic
textures of nanometric size \cite{martines2005,verho2012} proved efficient in
stabilizing the gaseous layer over a broad range of pressures and temperatures
\cite{checco2014}. However, the engineering of surface nanotextures is still in
its infancy:  understanding the collapse mechanism -- i.e., the path followed
by the liquid front during the breakdown of the Cassie state -- is the key to
improve the performance and robustness of such textures. For instance, it is
possible to modify surfaces in order to destabilize the Wenzel state
\cite{prakash2016} or to achieve robust superhydrophobicity via complex
nanotextures \cite{amabili2015}. Nucleation of gas and vapor bubbles, which is
enhanced by surface textures \cite{giacomello2013}, can also bring to the
breakdown of the Cassie state and superhydrophobic surfaces must be engineered
to prevent it.  However, to date the collapse mechanism on experimentally and
technologically relevant textures remains largely unknown, with the theoretical
approaches making \emph{a priori} assumptions or being often affected by
simulation artifacts as discussed in details below.

Due to their simplicity, surfaces decorated with pillars are a paradigm in the
study of superhydrophobicity, both via experiments and simulations.  In
addition, thanks to the reduced liquid-solid contact these kind of textures
favor the emergence of large slip at the walls \cite{ybert2007}, making them
attractive, e.g., for drag-reduction \cite{rothstein2010}.  The focus of this
work is the wetting of 3D submerged nanopillars (Fig.~\ref{fig:system}): the
three distinctive attributes of the textured studied in this work are
submerged, three-dimensional, and nanoscale.  Indeed, most simulations to date
have restricted their attention to drops of sizes comparable to that of the
texture at the top of which they are sitting
\cite{dupuis2005,kusumaatmaja2007,koishi2009,savoy2012,ren2014,zhang2014,pashos2015};
in this setup the collapse mechanism is significantly influenced by the drop
shape and size.  At variance with drops, the main issue for submerged surfaces
is the resistance of the superhydrophobic state against pressure variations
\cite{giacomello2012langmuir,savoy2012b,amabili2015,amabili2016,prakash2016}\footnote{A
further advantage of the submerged setup is that results are also relevant for
understanding the collapse of liquid drops with size much larger than the
nanotexture.}.  Present superhydrophobic surfaces generally exhibit complex 3D
morphologies with isolated (pores) or connected (pillars, cones) cavities.
Two-dimensional or quasi-2D (limited thickness) models have been often used in
simulations with the aim of reducing their computational cost
\cite{savoy2012,savoy2012b,giacomello2012langmuir,amabili2015,pashos2015}.
However, systems such as pillared surfaces cannot be reduced to 2D without
overlooking important phenomena, e.g.,  the simultaneous wetting of several
interpillar spacings.  Concerning the methods, mainly due to the significant
computational cost, most approaches dealing with 3D structures to date
considered a continuum description of the liquid (and solid)
\cite{ren2014,zhang2014,pashos2015,pashos2016,tretyakov2016,panter2016}, which
necessarily introduces some assumptions concerning the structure of the
interfaces, the interaction with the wall, and other nanoscale effects such as
fluctuations, line tension, etc. Such effects are in principle relevant on the
nanoscale.  In order to avoid these limitations and explore the fundamentals of
wetting of experimentally relevant, 3D nanotextures, here the collapse
mechanism of a large sample containing a $3$-by-$3$ array of pillars is
identified via molecular dynamics (MD) combined with the string method in
collective variables. Other atomistic approaches have been recently employed
which, however, lead to a discontinuous dynamics for the collapse of the
meniscus \cite{prakash2016} (a detailed analysis of this problem is reported
elsewhere \cite{giacomello2015}). 

In brief, the objective of this work is addressing two open questions: i)
finding the most probable collapse mechanism for submerged nanopillars and
computing the associated free energy profile; ii) validating simple continuum
models for nanoscale wetting, identifying which bulk, interface, and line terms
are relevant.

Two challenges are inherent to these problems: i) employ a model which captures
the physics at the nanoscale, including fluctuations, evaporation/condensation,
and line tension; ii) interpret the results in simple terms, which can be used
for developing design criteria.  In this work, a Lennard-Jones (LJ) fluid and
solid are used to tackle i). This approach is computationally expensive, but
includes all the relevant physics with minimal assumptions \cite{evans2015}.
Concerning ii), a sharp-interface continuum model is used to interpret the
atomistic results, which yields a simple and informative picture of the
phenomena and allows to shed light on the nanoscale effects.  Results show
that, differently from previous studies focusing on independent cavities
\cite{deconinck2011,giacomello2015}, the collapse mechanism for pillars is
characterized by the combination of local and collective effects. In
particular, the meniscus breaks the symmetry imposed by the confining geometry
and the collapse happens via the \emph{local} formation of a single liquid finger followed
by the correlated, collective wetting of the interpillar space of points far apart on the
surface.

In interconnected geometries the meniscus is expected to assume complex
morphologies during the collapse. In the present study the (atomistic) density
field $\rho(\bm x)$ is used in order to monitor the process: this choice allows
one to track any changes in the meniscus shape and, at the same time, is
straightforward to compute in atomistic simulations (Fig.~\ref{fig:system}).
The most probable wetting path is identified via the  string method by using
the density field to characterize  the system configuration. The identified
path corresponds to a sequence of density fields $\rho(\bm x)$  at discrete
steps along the collapse process.  The string method also allows for computing
the free energy along the most probable path $\Omega\left(\rho(\bm x)\right)$
and, thus, the free energy barrier determining the kinetics of the process.
Indeed, we show that under ambient pressure and temperature conditions the
wetting requires that the system climbs a sizable free energy barrier, which is
possible only thanks to \emph{thermal fluctuations} \cite{bonella2012,
Lelievre:FreeEnergyComputationsAMathematicalPerspective:2010}. 

The paper is organized in a methodological section, in which the main
conceptual  aspects of the string method are described, and a second,
self-contained section in which the results on the collapse of the
superhydrophobic state are discussed in depth.  Thus, the reader interested
only in the physical results can go directly to this latter section.  The final
section is left for conclusions.

%%%%%%%%%%%%%%%%%%%%%%%%%%%%%%%%%%%%%%%%%%%%%%%%%%%%%%%%%%%%%
\section{Methods}
%%%%%%%%%%%%%%%%%%%%%%%%%%%%%%%%%%%%%%%%%%%%%%%%%%%%%%%%%%%%%
\label{sec:methods}
Typically, large free energy barriers separate the
Cassie and Wenzel states. This means that, even if thermodynamics might
favor the process, i.e., the final is state more stable than the
initial one, the system still has to climb the free energy up to a
saddle point (barrier) for the wetting transition to take place. This is
different from barrierless processes; in presence of a barrier
wetting is possible only thanks to \emph{thermal fluctuations}, which
determine the timescale for observing a transition event. This
timescale is too long to be accessed by brute force MD: this is the
problem of \emph{rare events} \cite{bonella2012}.
To cope with this
issue, here the  string method in collective variables (CVs) is
used \cite{maragliano2006}. 
This method allows one to follow the infrequent process at fixed thermodynamics
conditions (i.e., without increasing pressure and temperature), which
implies visiting configurations characterized by high free energies and
a corresponding, exponentially low probability. 

CVs at the basis of the string method are a (restricted) set of observables $\{\theta_i(\bm{r})\}_{i=1,M}$
function of $\bm r$, the $3n_f$ dimensional vector of the fluid particles 
positions. CVs should correspond to the degrees of freedom which are
able to characterize the system along the transition from the initial to
the final state. 
Here we employ the coarse-grained density field, which   
has been used in recent works on wetting 
transitions \cite{giacomello2012langmuir,giacomello2015} and other hydrophobicity-related 
phenomena \cite{Miller:2007ws}. 
Moreover, the density field is the cornerstone of most of meso- and macroscopic
descriptions of homogeneous and heterogeneous fluids, e.g.,  the
\emph{classical} Density 
Functional Theory \cite{evans1992,lowen2002}. 
Thus, the coarse-grained density is a natural CV to study  
the collapse of the Cassie state. 

\begin{figure}
		\centering
		\includegraphics[width=0.48\textwidth]{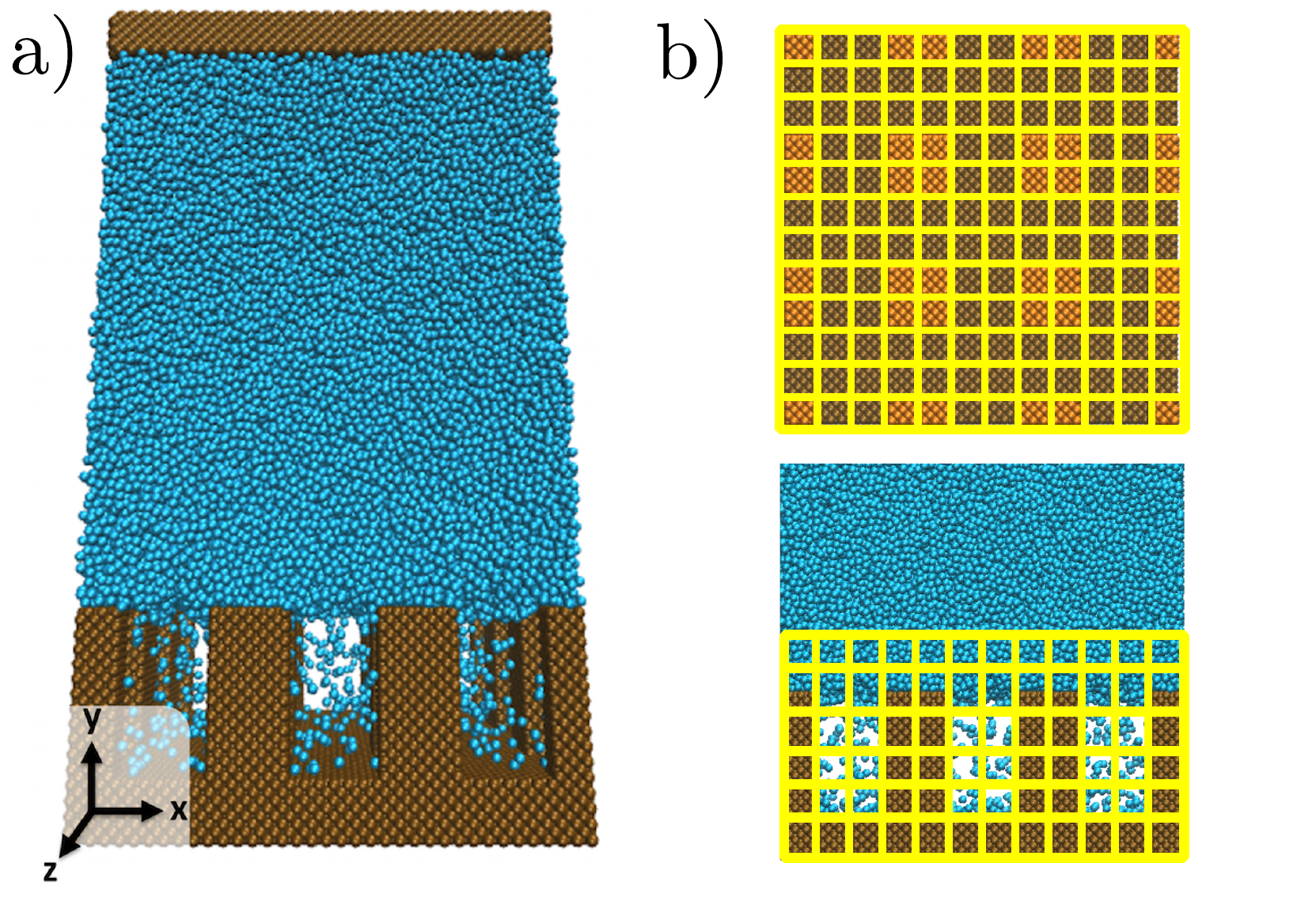}
		\caption{ a) Atomistic system, consisting of a LJ fluid (blue) and
			solid (brown). The upper solid wall serves as a piston to impose
			constant pressure; the lower one is decorated with $9$
			nanopillars. Temperature is controlled via a Nos\'e-Hoover chain
			thermostat \cite{martyna1992}.
			b) Definition of the collective variables used in the string
			method: coarse-grained density defined on $M=864$ cubic cells (in yellow; top and
			lateral views).
		\label{fig:system}}
		\end{figure}

The coarse-grained density field is  defined as 
\begin{equation}
	\theta_i(\bm{r}) \equiv \theta(\bm{r}, {\bm x}_i) = \frac{1}{\Delta V}
	\sum_{j=1}^{n_f} \tilde \chi(\bm{r}_j, 
  {\bm x}_i)
	\label{eq:CV}
\end{equation}  

\noindent where $\theta(\bm{r}, {\bm x}_i)$ is the density at the  point ${\bm x}_i$ of the 
discretized (or coarse-grained) ordinary space (Fig.~\ref{fig:system}b) and $\tilde \chi(\bm{r}_j, {\bm x}_i)$ 
is a smoothed approximation to the characteristic function of the volume $\Delta V$ 
around the point ${\bm x}_i$, with the sum at the RHS running over the 
$n_f$ fluid particles. In practice, these functions count the
number of atoms in each cell in Fig.~\ref{fig:system}b.
In the following, whenever possible, $\theta_i(\bm{r}) = \theta(\bm{r}, {\bm x}_i)$ is replaced by the 
lighter notation $\theta(\bm{r}, {\bm x})$, which denotes the CV at all points
$\bm{x}_i$ of the discretized space, and $\rho({\bm x})$ a realization of this CV.
Although $\theta(\bm{r}, {\bm x})$, and analogously  $\rho(\bm{x})$, is
interpreted as a coarse-grained density field, we find sometimes
convenient to consider it as an $M$-dimensional vector with components
$\theta_i(\bm{r})$ ($\rho_i$, respectively).

Our aim is to find and characterize the most probable path $\rho({\bm x}; \tau)$ 
followed by the system along the Cassie-Wenzel transition. This path represents a curve in the 
space of the coarse-grained density or, more visually, a series of snapshots
taken at successive times $\tau$ along the wetting process. The parametrization 
of this curve with the time $\tau$ of the transition is unpractical in
actual calculations \cite{e2007}. A more convenient but equivalent parametrization is 
the normalized arc-length parametrization, i.e., the parametrization according to the length of the 
curve $\rho(\bm{x}; \tau)$ at a given time $\tau$, divided by the total length of the curve: $\lambda = \int_{\rho_C}^{\rho(\tau)} 
\mathrm d\rho / \int_{\rho_C}^{\rho_W} \mathrm d\rho$, where $\rho_C$ and $\rho_W$ 
are the density fields corresponding to the Cassie and Wenzel states,
respectively, and $\mathrm d{\rho} = \sqrt{ \sum_{i=1,M} \left( \partial
\rho({\bm x}_i, \tau)/\partial \tau \right)^2} \mathrm d \tau$.

An infinite number of paths exists that bring the system from the Cassie
to the Wenzel state. The objective of the string method is to find the most probable one. 
 When the thermal energy $k_BT$ is low as compared to the free energy barrier the system has to overcome along the process,  this path must satisfy the 
 condition \cite{maragliano2006}  (see also the Appendix)
\begin{equation}
\Big [ \hat g\left (\rho({\bm x}; \lambda) \right)
\nabla_{\rho}\Omega\left (\rho({\bm x}; \lambda) \right) \Big ]_\perp = 0\, ,
\label{eq:MFEP}
\end{equation}
where $\nabla_\rho$ is the vector of derivatives with respect to the
different components of the collective variable $\rho$, namely, the vector whose components are 
$\partial/ \partial \rho_i$ with $i=1, \ldots, M$.
Here, the symbol $\mathbin{\perp}$ denotes ``orthogonal to the path $\rho({\bm x}; \lambda)$'';
$\Omega(\rho({\bm x}; \lambda))$ is the Landau free energy of the
coarse-grained density field $\rho({\bm x}; \lambda)$ along the string.
This free energy is in general related to the probability of observing the  realization $\rho({\bm x})$ of the coarse
grained density under the atomistic probability distribution $m({\bm r})$:
\begin{equation}
\Omega(\rho({\bm x}))\equiv - k_B T \ln p \left[ \rho({\bm x}) \right]=
-k_B T \ln \Big( \int \mathrm d\bm{r} \; m(\bm{r}) \prod_{i=1}^{M} 
\delta (\theta(\bm{r}, {\bm x}_i)-\rho({\bm x}_i)) \Big )
\; .
\label{landau}
\end{equation}
The $i$-$j$ components of the
metric tensor $\hat g(\rho({\bm x}))$ in Eq.~\eqref{eq:MFEP} read 
\begin{eqnarray}
g_{ij}(\rho({\bm x})) = \frac{\int \mathrm d\bm{r} \; \nabla_{\bm{r}} \theta(\bm{r},\bm x_i)  
\cdot \nabla_{\bm{r}} \theta(\bm{r},\bm x_j) \;  m(\bm{r}) \prod_{l=1}^{M} \delta
(\theta(\bm{r}, \bm{x}_l)-\rho(\bm{x}_l))}{\int\mathrm  d\bm{r} \; m(\bm{r})
\prod_{l=1}^{M} \delta
(\theta(\bm{r}, \bm{x}_l)-\rho(\bm{x}_l))} \, ,
\label{metric}
\end{eqnarray}
where $\nabla_{\bm{r}}$ denotes the $3n_f$-dimensional vector of the
derivatives with respect to the components of the particle positions, 
$\partial /\partial r^l_\alpha$, with $\alpha=1, \ldots, n_f$ and $l=x, y, z$.

The intuitive meaning of Eq.~\eqref{eq:MFEP} is that 
the component of the \emph{force} 
$\hat g\left (\rho({\bm x}; \lambda) \right)
\nabla_{\rho}\Omega\left (\rho({\bm x}; \lambda) \right)$ 
orthogonal to the path is zero. To give a visual interpretation, 
$\Omega(\rho(\bm x))$ should be imagined as a rough free energy landscape;  a path
satisfying condition~\eqref{eq:MFEP} lies at the bottom of a valley
connecting the initial and final states (free energy minima) and passes
through a mountain pass (free energy saddle point).

In atomistic simulations,
$\hat g\left (\rho({\bm x}; \lambda) \right)$ and 
$\nabla_{\rho}\Omega\left (\rho({\bm x}; \lambda) \right)$ can be
estimated, for each point $\lambda$ constituting the string, by an MD
governed by the following equations of motion \cite{TAMD}:
\begin{equation}
m_\alpha \ddot{\bm{r}}_\alpha = -\nabla_{\bm{r}_\alpha} V(\bm{r}) - 
k \sum_{i=1}^{M} (\theta(\bm{r}, {\bm x}_i)-\rho({\bm x}_i; \lambda))  
\nabla_{\bm{r}_\alpha} \theta(\bm{r}, {\bm x}_i) +
\text{thermo}(T) + \text{baro}(P) \text{ ,}
\label{restrained}
\end{equation}

\noindent where $m_\alpha$ is the mass of the $\alpha$-th particle,
$\alpha=1,\ldots,n_f$, $V(\bm r)$ is the interparticle potential, and $k \sum_{i=1}^{M}  (\theta(\bm{r}, {\bm x}_i)-\rho({\bm x}_i; \lambda)) 
\nabla_{\bm{r}_\alpha} \theta(\bm{r}, {\bm x}_i)$ is a biasing force
which allows the system to visit regions of the
phase space around the condition $\theta(\bm{r}, {\bm x}) = \rho({\bm x}; \lambda)$. The 
biasing force makes it possible to estimate the gradient of the free energy and the metric 
matrix of points having a very low probability (high free energy), such those near the transition state. 
In practice, $\hat g\left (\rho({\bm x}; \lambda) \right)$ and 
$\nabla_{\rho}\Omega\left (\rho({\bm x}; \lambda) \right)$ are computed as time 
averages along the MD of Eq.~\eqref{restrained}: 
\begin{eqnarray}
g_{ij}(\rho({\bm x}; \lambda)) &=& {1 \over t_{MD}} \int_0^{t_{MD}}
	\mathrm d s \; \nabla_{\bm{r}} \theta(\bm{r}(s),\bm x_i)  
	\cdot \nabla_{\bm{r}} \theta(\bm{r}(s),\bm x_j) \text{ ,}\nonumber \\
{\partial \Omega\left (\rho({\bm x}; \lambda) \right) \over \partial \rho_i} &=& - {1 \over t_{MD}} 
\int_0^{t_{MD}} \mathrm ds \;  k (\theta(\bm{r}(s), {\bm x}_i)-\rho({\bm
x}_i; \lambda)) \text{ ,} \nonumber
\end{eqnarray}
with $t_{MD}$ the duration of the MD simulation and the indices $i$ and $j$
running over the $M$ coarse-graining cells.

The (improved) string method \cite{e2007} is an iterative algorithm that, starting from a first-guess path, 
produces the most probable path $\rho({\bm x}; \lambda)$ satisfying 
Eq.~\eqref{eq:MFEP}. Here we refrain to describe the algorithm in detail
summarizing only the main steps of the method. For further details the interested reader is
referred to the Appendix, to the original
article \cite{e2007}, or to reviews \cite{bonella2012}.  
In the string method, the continuum path $\rho({\bm x}; \lambda)$, with
$0\leq \lambda\leq 1$, is replaced by its discrete counterpart
$\{\rho({\bm x}; \lambda_n)\}_{n=1,L}$, with $L$ the number of
snapshots used to discretize the path. The algorithm starts from a first guess of the 
wetting path, $\{\rho^0({\bm x}; \lambda_n)\}_{n=1,L}$, and performs an iterative 
\emph{minimization procedure} which yields a path with zero orthogonal 
component of the force: $\{[\hat g\left (\rho({\bm x}; \lambda_n) \right)
\nabla_{\rho}\Omega\left (\rho({\bm x}; \lambda_n) \right)]_\perp =0 \}_{n=1,L}
$.
At each iteration, $\{\hat g\left (\rho({\bm x}; \lambda_n) \right)\}_{n=1,L}$ and 
$\{\nabla_{\rho}\Omega\left (\rho({\bm x}; \lambda_n) \right)\}_{n=1,L}$  are computed by  
the biased MD in Eq.~\eqref{restrained} and are used to generate a new path. 
The procedure is performed until convergence is reached, i.e., when the
difference in the free energy of each image between two string
iterations is below a prescribed threshold  (see Supplemental Material for convergence plots \footnote{see Supplemental Material at
[URL will be inserted by publisher] for simulation details}).

Along the path of maximum probability one can compute the free energy profile 
and the associated barrier, which can in turn be related to the wetting and 
(dewetting) transition time $t_{CW}$ ($t_{WC}$) \emph{via} the transition state theory: $t_{CW} = t^0_{CW} \exp[\Omega^\dag_{CW}/k_BT]$ ($t_{WC} = t^0_{WC} \exp[\Omega^\dag_{WC}/k_BT]$). In the present 
article, the free energy profile along the most probable wetting path is 
reported against the \emph{filling fraction} $\Phi(\lambda_n) =
(N(\lambda_n) - N_C) / (N_W - N_C)$, where 
$N(\lambda_n) \equiv \Delta V \sum_i^M \rho(\bm x_i;\lambda_n)$ is the number of liquid particles in the yellow region of
Fig.~\ref{fig:system}b at the string point $\lambda_n$ and $N_C$ and $N_W$ are the number of particles in the Cassie and Wenzel states, respectively.  
In the present calculations,
$N$ and $\Phi$ are non-decreasing functions of $\lambda$ along the wetting path; thus, the 
filling fraction can  be used as an \emph{a posteriori} parametrization of the most
probable wetting path $\rho({\bm x}; \Phi)$. It is important to remark that this does not imply that $\Phi$ 
is a good CV for the wetting transition in interconnected textured surfaces of the type studied in this
article. In fact, as shortly discussed in the next section,
and more extensively in a forthcoming article, using $\Phi$ or $\theta$ as 
CVs leads to qualitatively different results, with those of the first CV
characterized by an unphysically discontinuous path.

		%Fig1
%%%%%%%%%%%%%%%%%%%%%%%%%%%%%%%%%%%%%%%%%%%%%%%%%%%%%%%%%%%%%
%String vs other methods
Summarizing, the aim of the string method in CVs is to compute the most
probable path in a complex high dimensional free energy landscape. This
is different from the objective and approach of other techniques (e.g.,
umbrella sampling, restrained MD, boxed MD, etc.), which
require to reconstruct the entire landscape within a predefined volume of the
CV space. The advantage of the string method is that it scales linearly
with the number of CVs, while more standard methods scale exponentially.
This allowed us to investigate the collapse mechanism in the high
dimensional space (864 degrees of freedom) of the coarse grained density
field, which would have been impossible with other approaches.

%%%%%%%%%%%%%%%%%%%%%%%%%%%%%%%%%%%%%%%%%%%%%%%%%%%%%%%%%%%%%
%METHODS
MD simulations for computing $\{\hat g\left (\rho({\bm x}; \lambda_n)
\right)\}_{n=1,L}$ and $\{\nabla_{\rho}\Omega\left (\rho({\bm x}; \lambda_n)
\right)\}_{n=1,L}$ are performed at constant pressure, temperature, and number
of particles \cite{Note2}.  The
LJ fluid is characterized by a potential well of depth $\varepsilon$ and a
length scale $\sigma$. The (modified) attractive term of the LJ potential
between the fluid and solid particles is scaled by a constant to yield a Young
contact angle \emph{on a flat surface} $\theta_Y=113^\circ$.  The pressure is
close to the liquid-vapor bulk coexistence, $\Delta P \equiv P_l -P_v \simeq
0$, while the temperature is $T=0.8\,\varepsilon k_B^{-1} $. Free energy
profiles at different pressures can be obtained by simply adding an analytical
bulk term of the form $(P_l -P_v)\,V_v$, with $V_v$ the volume of the vapor
bubble \cite{Giacomello2012,prakash2016,panter2016,amabili2016,Amabili2016a}
which can be obtained form the number of fluid particles in the cavity region
(the yellow framework in Fig.~\ref{fig:system}).  The first guess for the
string calculation is obtained by running an MD simulation at high pressure. In
this condition the wetting barrier is small enough that the transition  occurs
on the timescale accessible by brute force MD. We remark that the rest of
simulations for the iterative string calculations are not run at such high
pressure but close to coexistence, as said above.  The simulations are run via
a modified version of the LAMMPS software \cite{LAMMPS}.

%%%%%%%%%%%%%%%%%%%%%%%%%%%%%%%%%%%%%%%%%%%%%%%%%%%%%%%%%%%%%
\section{Results and discussion}
\label{sec:ResultsAndDiscussion}
%%%%%%%%%%%%%%%%%%%%%%%%%%%%%%%%%%%%%%%%%%%%%%%%%%%%%%%%%%%%%

%%%%%%%%%%%%%%%%%%%%%%%%%%%%%%%%%%%%%%%%%%%%%%%%%%%%%%%%%%%%%
\subsection{Mechanism of collapse}

		%Fig2
		\begin{figure*}
		\centering
		\includegraphics[width=0.9\textwidth]{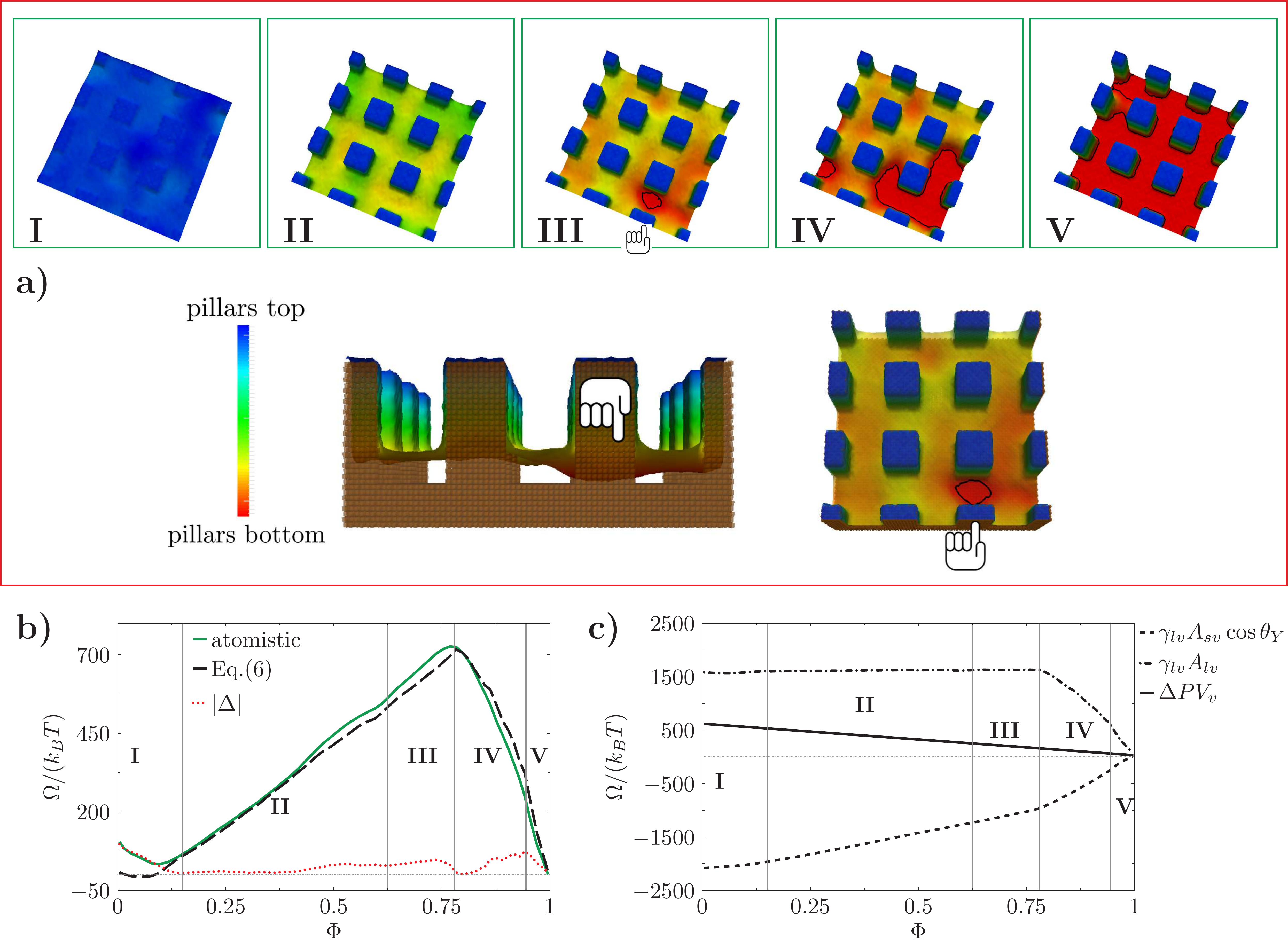}
		\caption{a) Collapse mechanism of the superhydrophobic Cassie
			state, shown by the density isosurface $\rho^\ast=0.5\, \rho_l$, where
			$\rho_l$ is the bulk liquid density at the simulated pressure. 
			Images I--V illustrate typical interface shapes corresponding to
			the different regimes discussed in the text. Black isolines
			identify the region where the liquid is in contact with the bottom wall
			(III--IV); color code indicates the meniscus  elevation over the
			pillars bottom; the hand symbol identifies the liquid finger.
			The bottom images show the meniscus shape at the transition state,
			where a liquid finger is formed.  			
			b) Free energy profile as a function of the filling fraction 
			at $\Delta P = 0.011\,\varepsilon\sigma^{-3}$ as
			computed via the string method in collective variables (green) and via the macroscopic
			theory in Eq.~\eqref{eq:macro} (black). The absolute value of the
			deviation between the two estimates is shown in red dotted line.
			c) Continuum analysis of the atomistic results based on
			Eq.~\eqref{eq:macro}: the solid-vapor, liquid-vapor, and volume
			contributions to the free energy are shown in dashed, dot-dashed,
			and solid lines, respectively. The volume of vapor is
			related to the vapor filling via $V_v=\Phi(N_C-N_W)/(\rho_l-\rho_v)$. 
		\label{fig:mechanism}}
		\end{figure*}

% String in summary %%%%%%%%%%%%%%%%%%%%%%%%%%%%%%%%%%%%%%%%%%%%
The Cassie-Wenzel transition can take place following different wetting
paths; here we analyze the one having the highest probability.  
At the pressures and temperatures of interest for technological
applications (e.g., drag reduction), the Cassie and Wenzel states are
separated by large free energy barriers that the system has to overcome.
Thus, even if thermodynamics might favor the collapse, the system still
has to climb the free energy barrier separating the initial and final
state. This apparently non spontaneous process can take place only
thanks to thermal fluctuations \cite{bonella2012,
Lelievre:FreeEnergyComputationsAMathematicalPerspective:2010}. This
implies infrequent transition events or, which is equivalent, long
transition times, much longer than the typical timescale accessible to
brute force (standard) MD.
Thus, here the most probable wetting path is
computed using a special technique  -- the string method in collective
variables \cite{maragliano2006}, which is described in detail in
Sec.~\ref{sec:methods}. It is important to remark that in the string
method the pressure and temperature are kept constant at the prescribed
values; in other words, the path obtained via the string method is
representative of the transition at constant pressure and temperature.

The path is constituted by a sequence of atomistic coarse-grained
density fields $\rho(\bm x)$ (Eq.~\eqref{eq:CV}), i.e., the path can be
seen as a series of snapshots of the density field taken at different
times along the transition from the Cassie to the Wenzel state. It is
important to remark that each $\rho(\bm x)$ is obtained by performing
the ensemble average over atomistic configurations consistent with the 
state of the system along the path; thus, $\rho(\bm x)$ contains all the 
nanoscale information.
 The morphology of the interface between the liquid and the vapor
 domains, the meniscus, is a simpler and more visual observable to
 follow the collapse mechanism. Thus, in the following the meniscus,
 which can be computed from $\rho(\bm x)$, will be used to describe the
 change of morphologies of the liquid along the collapse process, and
 its effect on the free energy. These morphologies will be discussed in
 detail, with particular attention to clarify the role of nanoscale
 aspects such as line tension.

Figure~\ref{fig:mechanism} shows the collapse mechanism of the
superhydrophobic state and the related free energy profile
$\Omega(\Phi)$ along the most probable wetting path at a pressure and
temperature close to the Cassie-Wenzel coexistence (i.e., where the
Cassie and Wenzel states have approximatively the same free energy, $T=0.8\,\varepsilon
k_B^{-1} $ and $\Delta P = 0.011\,\varepsilon\sigma^{-3}$). $\Phi$ denotes the
filling fraction defined as $\Phi \equiv (N - N_C) / (N_W - N_C)$, with
$N$ the total number of liquid particles in the yellow regions of
Fig.~\ref{fig:system}b and $N_C$ and $N_W$ the number of particles in
the Cassie and Wenzel
states, respectively. 
Here and in the 
following, notations of the type  $\Omega(\Phi)$ are used as a shorthand for 
the more complete notation $\Omega\left (\rho({\bm x}; \Phi) \right)$.
The wetting path is divided into five parts (I--V), corresponding to different regimes. 

In I, the meniscus is pinned at the top corners of the pillars, with
its curvature increasing with $\Phi$. The free energy minimum
corresponds to the Cassie state characterized by an almost flat
meniscus as expected at low $\Delta P$.

In II, the meniscus depins from the corners and progressively slides
along the pillars filling the interpillar space. The liquid-vapor
interface has a small curvature which
is sufficient to meet the pillars with the Young contact angle. 
Configurations in II correspond to a linear
increase of the free energy. Given the large ratio between the height of
the pillars and the spacing among them, the ``sag''
mechanism \cite{patankar2010,deconinck2011}, in which it is assumed that
the meniscus remains pinned at the pillars corner along the entire
process, cannot be realized.

The transition state (TS, free energy maximum) is found in III and
corresponds to the liquid touching the bottom wall. In this part of the
transition, the shape of the liquid-vapor interface changes
dramatically: starting from an almost flat interface a liquid finger forms 
between two pillars and then touches the bottom wall 
(see Fig.~\ref{fig:mechanism}a and \footnote{see Supplemental Material
at [URL will be inserted by publisher] for a video of the wetting transition}). 
The free energy barrier for collapse is very high for the considered
pressure $\Delta P = 0.011\,\varepsilon\sigma^{-3}$, $\Delta
\Omega^\dag_{CW}=670\;k_BT$. From this contact point, the liquid progressively
fills the surrounding interpillar spaces. In the regions in which the
liquid touches the bottom, the double liquid-vapor/vapor-solid interface
is replaced by the  liquid-solid interface, causing the free energy to
decrease. It is clear that identifying correctly the most probable TS is
especially important because this configuration determines the free energy barrier and,
with exponential sensitivity, the collapse kinetics. In
Sec.~\ref{sec:bib} a comparison
with other mechanisms proposed in the literature \cite{patankar2010,prakash2016}
shows how seemingly small morphological changes are reflected in
radically different estimates of these quantities.

		%Fig3
\begin{figure}[tb!]
		\centering
		\includegraphics[width=0.48\textwidth]{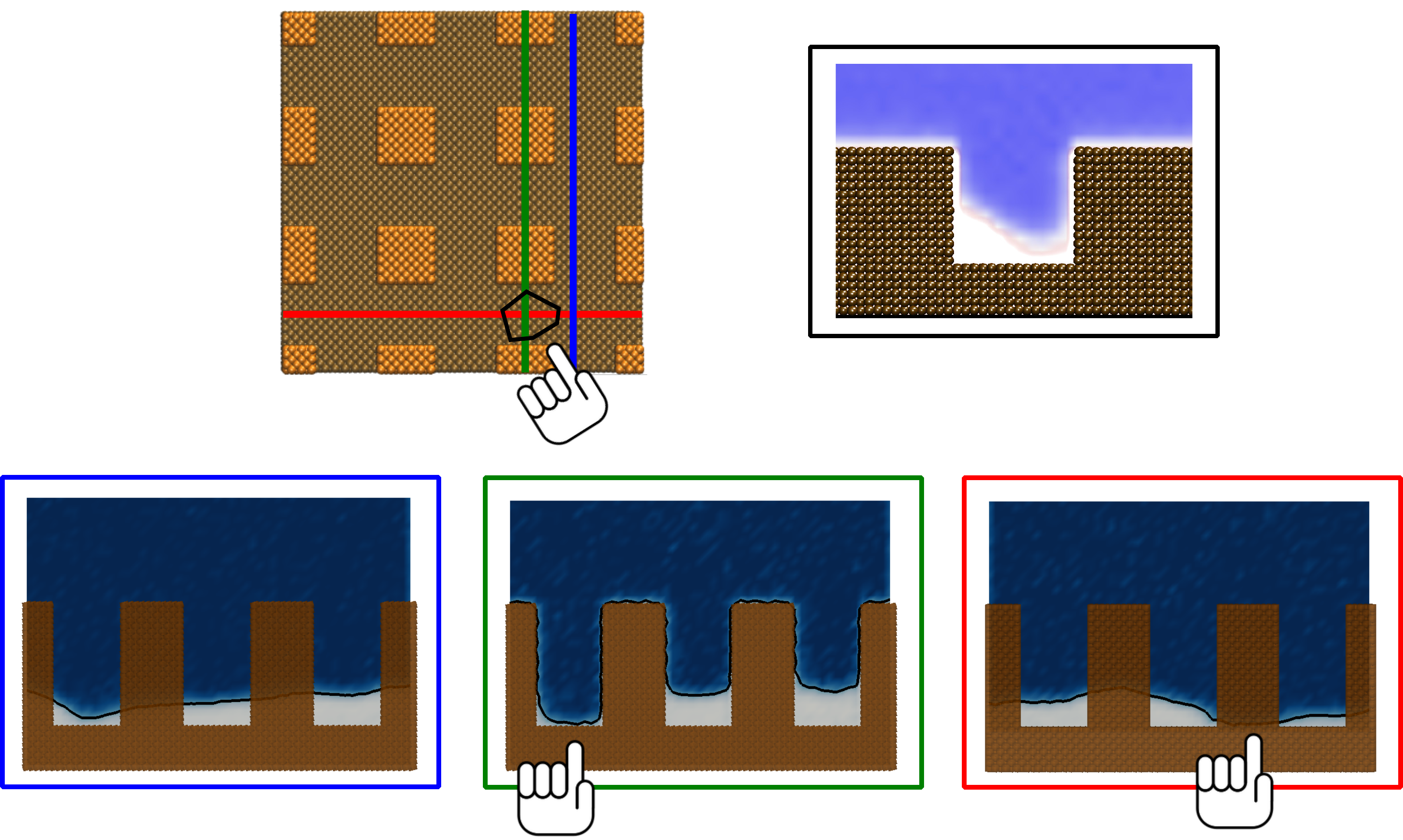}
		\caption{Two-dimensional cuts of the number density at the transition state: the liquid
			finger (identified by the hand symbol) forms at the intersection of the red and green planes.
			Bulk liquid density is represented in blue and vapor in white; black isolines identify
			the liquid-vapor interface. The density field at the top right is
			taken from the string simulations of Ref. \cite{giacomello2015},
			where a quasi-2D groove is considered.
		\label{fig:TS}}
\end{figure}

A closer look to the configurations around the TS shows that the finger
formation involves both quasi-2D and collective wetting of the pillars
(Fig.~\ref{fig:TS}). In one direction (green line in
Fig.~\ref{fig:TS}), the liquid finger is strongly
confined between pillars. Confinement renders the wetting mechanism
effectively 2D close to the pillars centers, as demonstrated by
comparing the second cut of the density field in Fig.~\ref{fig:TS} with the 2D
mechanisms of previous works \cite{giacomello2015}: 
the wetting of the bottom wall happens via
the formation of two asymmetric bubbles at the pillars lower corners
with the smaller bubble disappearing faster. In the orthogonal direction
(red line in
Fig.~\ref{fig:TS}) and far from the confined finger (blue line)
the liquid-vapor interface must bend in order to recover the flat shape; 
the ensuing curvature of the interface is gentle because of surface
tension. This collective, large scale mechanism triggers the wetting of the
interpillar spaces surrounding the initial finger, causing the final
collapse of the vapor domain.

Figures~\ref{fig:mechanism}a and \ref{fig:TS} show that the
collapse is asymmetric and involves many interpillars spacings; this
fact suggests that assuming a symmetric mechanism \cite{patankar2010} or simulating a single elementary cell is insufficient
to capture correctly the mechanism and the related free energy since it
imposes an unphysical symmetry to the problem. Even for the large domain
simulated here, the effect of periodic boundary conditions becomes
apparent at large filling levels beyond the TS; this however does not
affect the estimation of the free energy barriers.

		%Fig4
\begin{figure}[tb!]
		\centering
		\includegraphics[width=0.48\textwidth]{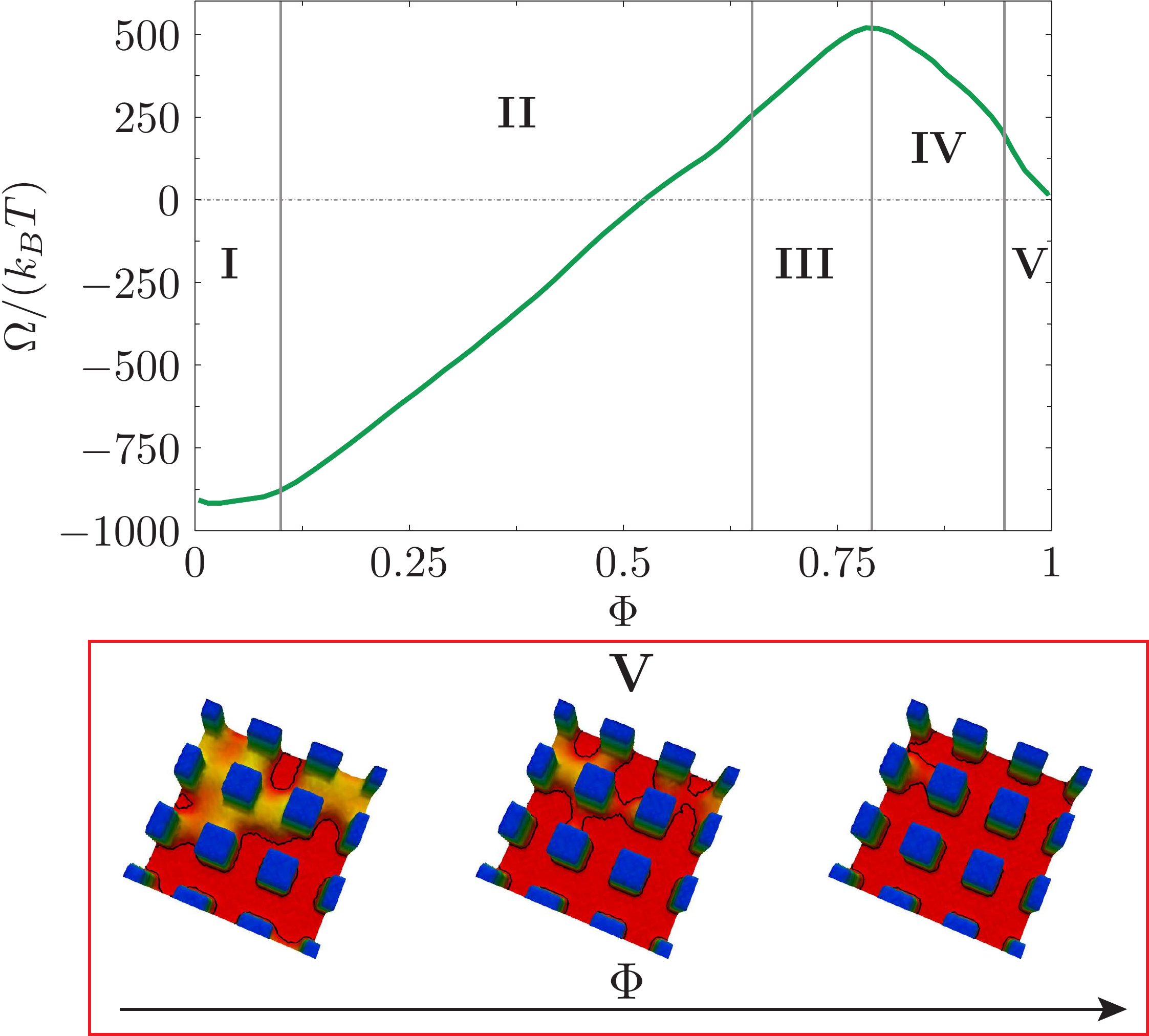}
		\caption{Top)  Free energy profile at bulk liquid-vapor coexistence,
			$\Delta P\approx 0$ computed via the atomistic string.
			Bottom) Formation of density-depleted regions in between the
			pillars in part V of the collapse and percolation thereof. Same color code as in
			Fig.~\ref{fig:mechanism}.
			\label{fig:depletion}}
\end{figure}

The present simulations can also help understanding the mechanism of the
opposite process, i.e., the dewetting transition from Wenzel to Cassie.
Indeed, under the hypothesis of a quasi-static transformation implicit
in the string method, the forward and backward processes happen
reversibly along the same path \footnote{The quasi-static assumption also implies that
the effect of viscosity and other kinetic effects are neglected in the
present approach.}.
Thus, here we describe the Wenzel-Cassie process associated to the path in Fig.~\ref{fig:mechanism} paying particular attention to the aspects
relevant to dewetting. In order to do that in Fig.~\ref{fig:depletion}
we report the free energy profile at conditions close to bulk liquid
vapor coexistence, $\Delta P\approx 0$. Dewetting, corresponding
to the nucleation of a vapor domain in the textured surface, starts with
the formation of two low-density, flat domains at the bottom of two
facing pillars. These domains grow and merge, forming a single depleted
region in between the pillars.  This domain then percolates to
the neighboring pillars, forming a connected network.  While the string
identifies only one percolating network, the texture geometry suggests
that many such
networks are possible, all of which involve multiple pillars. It must be stressed that
this initial ``collective'' nucleation process, which corresponds to part V of
Fig.~\ref{fig:mechanism}, cannot be captured by simulating a single pillar or an
elementary cell in one or two directions \cite{prakash2016,panter2016,savoy2012,savoy2012b}. 
Further moving toward the Cassie state, the percolating, depleted domain starts to
form also in the spaces among four pillars, and detaches from the
bottom forming a proper bubble. This part of the path corresponds to domain IV, where the free
energy shows a descending trend with different slope. This branch of the process is completed
when the liquid finger finally detaches from the bottom of the surface
in a point in between two pillars, domain III.
The free energy barrier for dewetting at $\Delta P\approx 0$ is $480\;k_BT$.

%%%%%%%%%%%%%%%%%%%%%%%%%%%%%%%%%%%%%%%%%%%%%%%%%%%%%%%%%%%%%
\subsection{A sharp-interface interpretation of the atomistic mechanism}

In order to rationalize the atomistic results in simple energetic terms, 
a macroscopic, sharp-interface model of capillarity is used in
connection with the data summarized in Fig.~\ref{fig:mechanism}.
For this model the free energy reads \cite{kelton2010}:
\begin{equation}
	\Omega_\mathrm{macro} \equiv \Omega - \Omega_\mathrm{ref} = \Delta P V_v +  
	\gamma_{lv} \left( A_{lv} +	\cos \theta_Y A_{sv} \right) \text{ ,}
	\label{eq:macro}
\end{equation}
where the Wenzel free energy $\Omega_\mathrm{ref}=\Delta P V_\mathrm{tot} -
\gamma_{lv} \cos\theta_Y A_\mathrm{tot}$ is taken as a reference and
$V_\mathrm{tot}$ and $A_\mathrm{tot}$ are the total volume of the interpillar space
and its internal surface area, respectively. 
It is important to remark that $\Omega_\mathrm{macro}$ actually depends on the shape of the liquid-vapor
interface $\Sigma_{lv}$, which, in turn, determines the solid-vapor one
$\Sigma_{sv}$ and the vapor volume $V_v$.
Since $\Omega_\mathrm{macro}$ defined in  Eq.~\eqref{eq:macro} is a
functional of the complete $\Sigma_{lv}$, it can be computed for arbitrarily complex
configurations of the capillary system, e.g., along wetting paths which may or may not contain collective effects. 
The first term on the RHS corresponds to the bulk energy of
a system containing a vapor bubble of volume $V_v$: at $\Delta P>0$ vapor
bubbles are energetically hindered while at negative pressures they are
favored. This term is the driving force for the liquid-vapor transition. 
The second terms on the RHS are the energy cost of the
liquid-vapor and solid-vapor interfaces, respectively; both are
multiplied by the liquid-vapor surface tension $\gamma_{lv}$.  The cost of $A_{sv}$ is
modulated by the Young contact angle $\cos\theta_Y\equiv (\gamma_{sv} -
\gamma_{sl})/\gamma_{lv}$.
In the hydrophobic case
considered here ($\theta_Y>\pi/2$), it is energetically favorable to
increase $A_{sv}$, e.g., by dewetting the pillars: this is why at the liquid-vapor bulk coexistence ($\Delta P\approx 0$) the Cassie state is thermodynamically stable.

The terms in Eq.~\eqref{eq:macro}, which are needed in order to make a
connection with the atomistic simulations, are consistent with the
properties of the atomistic system and can be computed as follows (see
Supplemental Material for details on the calculations of
the contact angle, surface tension, and areas \cite{Note2}).
$\Delta P=0.011\,\varepsilon \sigma^{-3}$ is controlled
by the barostat, while $\gamma_{lv}=0.57\,\varepsilon \sigma^{-2}$ 
and $\theta_Y=113^\circ$ are computed via independent MD simulations. 
The liquid-vapor and solid-liquid areas are computed by applying the marching cube method \cite{Lorensen:1998de}
on the $1\times 1\times 1\,\sigma^3$ coarse-grained density field obtained from the string
calculations. The volume of vapor is computed as  $V_v\equiv
(N_W-N)/(\rho_l-\rho_v)$, where $N_W$ and $N$ are the number of
particles within the yellow region of Fig.~\ref{fig:system} in the
Wenzel and in a generic state, respectively, and $\rho_l$ and $\rho_v$ are the bulk liquid and vapor
densities.  It is
important to remark that all the physical quantities are computed
independently, i.e., they are not fitted to reproduce the atomistic free energy profile of wetting. 

Figure~\ref{fig:mechanism}b shows a fair agreement between the
atomistic free energy and Eq.~\eqref{eq:macro}. 
In particular, the model in Eq.~\eqref{eq:macro} \emph{does not} include line tension, which is often
invoked to explain in macroscopic terms nanoscale wetting
phenomena \cite{sharma2012,guillemot2012}. One notices that the energetic
difference between the atomistic and macroscopic models (dotted red
curve of Fig. \ref{fig:mechanism}b) is a small fraction of the wetting
barrier, $\approx 10~\%$. This suggest that nanoscale effects, which
include line tension, dependence of the surface tension on the curvature
of the meniscus, width of the interfaces, etc., play a minor role in the
transition path and kinetics of the wetting. In particular, considering that
the contact line changes significantly in the regimes III-V, the present
results indicate that line tension does not change dramatically the free
energy profile for hydrophobic pillars on the $3$~nm scale, pushing
further down the limit where line effects might be relevant
(recent experimental work on superhydrophobicity \cite{checco2014} showed the same
for textures down to $10$~nm).
More precisely, the combination of atomistic simulations with the
continuum analysis entailed in Eq.~\eqref{eq:macro} shows
that the contribution of line tension to the intrusion or
nucleation free energy barriers is negligible for the present system. 
It will be interesting to use a similar approach to explore extreme
confinement (of the order of a nanometer or less), where a major
contribution of line tension is expected \cite{sharma2012}.
The differences observed in the pinning region might be ascribed to the
increased thickness of the liquid-vapor interface. In terms of
macroscopic theories, the latter are associated to finite temperature
effects, such as capillary waves. These effects, neglected in
Eq.~\eqref{eq:macro}, are sizable at depinning and seem to make the
sharp-interface model less reliable there.

Apart from these minor limitations,  Eq.~\eqref{eq:macro} provides a simple interpretation of
the atomistic collapse mechanism shown in Fig.~\ref{fig:mechanism}a. 
In I, the area of the liquid-vapor interface $A_{lv}$ slightly
increases, reflecting the increasing curvature due to pinning. At the same time the top surface 
and corners of the pillars become wet and the area of the solid-vapor interface $A_{sv}$
decreases.  

In II, $A_{lv}$ remains constant, while $A_{sv}$ linearly decreases,
which corresponds to an increase of the interface area between the
liquid and the hydrophobic solid. This surface term more than balances
the decreasing bulk term $\Delta P V_v$, resulting into a linear increase of the free energy. If the interface were to keep this ``flat'' configuration until the bottom
surface is fully wet, as assumed in classical
works \cite{patankar2003,patankar2010}, the linear trend would
continue, determining a much larger free energy barrier $\Delta
\Omega_{CW}^\dag=920\;k_BT$. This figure would lead, in turn, to
estimate a lifetime of superhydrophobicity exponentially longer than for
the mechanism shown in Fig.~\ref{fig:mechanism} ($10^{108}$ times).
Instead, at variance with the classical mechanism, at a distance $h_{TS}\approx 6\,\sigma$ from the bottom surface (part III), it
is  energetically more convenient to bend the liquid-vapor interface
between two pillars. The free energy of formation of this finger is seen
to be negligible, while it allows the liquid to wet the bottom surface.
Thus, the two  liquid-vapor and vapor-solid interfaces are replaced by a single liquid-solid one, which decreases the total free energy by 
$\Omega_\mathrm{dry}-\Omega_\mathrm{wet}=\gamma_{lv}(A_\mathrm{lv,dry} + \cos\theta_Y
A_\mathrm{sv,dry})>0$ (Fig.~\ref{fig:mechanism}c).

%%%%%%%%%%%%%%%%%%%%%%%%%%%%%%%%%%%%%%%%%%%%%%%%%%%%%%%%%%%%%
\subsection{Comparison with results in the literature}
\label{sec:bib}

It is now possible to compare the path in Fig.~\ref{fig:mechanism}
obtained via the atomistic string with those available in the literature
and computed via continuum \cite{ren2014,pashos2016} and atomistic \cite{savoy2012,prakash2016} simulations. 
Pashos \emph{et al.} \cite{pashos2016} considered a sharp interface model
for the liquid-vapor interface with an empirical diffuse liquid-solid
potential, which was introduced in order to simplify the numerical treatment of
complex textures and topological changes of the meniscus (for a similar
approach, see Ref. \cite{giacomello2015} -- 
see also Supplemental Material for the effective diffuse liquid-solid potential
computed via MD for the present solid-liquid interactions \cite{Note2}). 
The pillar height-to-width
ratio is very low, such that the surface is physically reminiscent of a
naturally rough one rather than the artificial textures used  for
superhydrophobicity \cite{checco2014}. In particular, the interaction length of the  liquid-solid potential is roughly
half the pillar height. The path obtained with this model is still
characterized by the formation of a	liquid finger touching the bottom wall in a single
point with the liquid front then moving to neighboring pillars. However,
the free energy profile is very different from the present, with a
number of intermediate transition states. This discrepancy is not
unexpected, given the different pillars height and the different range of
potentials employed.

Ren and coworkers \cite{ren2014,zhang2014} combined the string method
with a mesoscale diffuse interface model in order to study the
impalement of a drop on 3D surfaces with (slender) pillars. Although the
system is rather different from the present one, the finger mechanism
was found to be always energetically favorable. The finite lateral
extent of the drop, however, significantly alters the advanced phase of
the Cassie-Wenzel transition.

Savoy and Escobedo \cite{savoy2012} studied the Cassie-Wenzel transition
of a nanodroplet on quasi-2D pillars. In their  atomistic
trajectories obtained via forward flux sampling (FFS) \cite{allen2009}
the formation of a liquid finger can also be identified in some
atomistic configurations. However it is difficult to compare other
details of the path and the energetics, because the FFS method only
gives access to independent atomistic trajectories and not to the most
probable density field as for the string.

Patel and coworkers \cite{prakash2016} used umbrella sampling to study
the collapse of the submerged superhydrophobic state on pillars. This
geometry is very close to the present one, the main difference being
that there a single elementary cell was studied, which, due to the
periodic boundary conditions, corresponds to the wetting of a single
pillar. While a sort of finger formation is observed in
correspondence of the transition state, the subsequent wetting process
substantially differs from that in Fig.~\ref{fig:mechanism} in two
respects. The first one is that the wetting  process is discontinuous, with jumps in the density
field between neighboring configurations. This is a known artifact
ensuing from the use of a single collective variable
\cite{giacomello2015} that cannot distinguish among bubble shapes
enclosing the same volume $V_v$; a detailed analysis on the choice of
the CVs will  be object of an upcoming work.
The second
difference is the shape of the vapor bubbles during the later stage of
the transition: the use of a single elementary cell imposes a symmetry
to them which does not necessarily correspond to the most probable one.

Summarizing this analysis of available simulations, the formation
of a liquid finger has been reported in a number of previous works
dealing with both drops and submerged surfaces, tall and short posts
\cite{dupuis2005,kusumaatmaja2007,savoy2012,ren2014,pashos2016,prakash2016,panter2016}.  
There are, however, also relevant differences in the configuration of this transition state.
Here, the liquid finger forms between two pillars; in such a way the
liquid-vapor interface bends only on two sides of the pillars, while the
remaining sides become wet. This is energetically favorable because the
cost of replacing a portion of the two liquid-vapor and  solid-vapor interfaces with a solid-liquid interface is lower than 
that of increasing $A_{lv}$, $-\gamma_{lv}\cos\theta_Y<\gamma_{lv}$. 
Therefore it seems that the exact point where the finger forms and, consequently, where the transition state occurs,
depends on the details of the confining geometry (in particular on the height/spacing ratio of the pillars) and on the chemistry of the surface. 

In general terms, the presence of the bottom surface breaks the
translational symmetry of the meniscus sliding along the
pillars and thus imposes a change of topology to the vapor bubble. The
\emph{local} formation of a single, relatively narrow liquid finger is
the energetically favored way to accomplish this transition. 
It is still an open question whether the finger should touch the bottom surface symmetrically or not, 
i.e., if two equivalent vapor bubbles are formed at the bottom of the two pillars confining the finger. 
For the 2D groove geometry, simulations have suggested that the asymmetric
pathway is energetically favored because it allows the formation of a
single vapor bubble in a corner \cite{Giacomello2012,giacomello2015}.
However, this finding does not exclude that other pathways are possible
and might also be favored for other reasons (e.g., kinetic/inertia reasons).
Indeed recent experiments \cite{lv2014b} on a similar geometry suggest
that both pathways are possible. 
Moreover, recent string calculations \cite{panter2016} underscore that in 2D grooves  pathways  with both symmetric and asymmetric bubbles are possible, with the latter having a
slightly lower free energy barrier. 

On pillared surfaces, careful analysis of the collapse of the
superhydrophobic state by confocal microscopy \cite{papadopoulos2013} revealed the (abrupt) formation of vapor bubbles at
one of the pillars' lower corners, after the first contact with the
bottom wall. This asymmetry in the last stage of the wetting process
possibly indicates that the finger formation, which is too fast to be
experimentally observed, was asymmetric too.

%%%%%%%%%%%%%%%%%%%%%%%%%%%%%%%%%%%%%%%%%%%%%%%%%%%%%%%%%%%%%
\section{Conclusions}
%%%%%%%%%%%%%%%%%%%%%%%%%%%%%%%%%%%%%%%%%%%%%%%%%%%%%%%%%%%%%

In summary, rare-event atomistic simulations have provided a detailed
description of the collapse of the superhydrophobic Cassie state on a
nanopillared surface. This system is often encountered in experiments
and serves as a prototype for wetting of interconnected cavities.
Results have showed that the collapse proceeds in five stages: depinning
of the liquid-vapor interface from the pillars' corners; progressive
wetting of the pillars via an almost flat, horizontal meniscus;
formation of a liquid finger between two pillars which touches the
bottom wall; progressive filling by a non-flat meniscus; absorption of
a percolating network of low-density domains connecting pairs of pillars. 
This mechanism is very different from 2D grooves
\cite{patankar2003,deconinck2011,giacomello2015} and also from other 3D
simulations \cite{prakash2016} and could be captured only with state-of-the-art
simulation techniques. On one hand, results indicate that the system
should be large enough to capture the local formation of a liquid
finger and the progressive filling of the surrounding spaces. 
On the other hand, the number and choice of collective variables should
allow one to resolve density changes in between the pillars. 
These caveats should be born in mind for simulations of wetting:
artifacts can conceal the actual collapse mechanism, affect the estimation
of the free energy barriers and, with exponential sensitivity, of the
collapse kinetics.
The continuum analysis of the MD results entailed in
Eq.~\eqref{eq:macro} has allowed us to identify the different area and
volume contributions to the free energy along the collapse, showing that
line tension, evaporation/condensation, and other nanoscale effects
are negligible in the present conditions. 
These results shed new light on the wetting process of interconnected
gas reservoirs which can open new perspectives in the design of nanotextures.

\appendix
\section{Appendix: the string method in collective variables}

Referring to the specialized literature for a complete derivation
\cite{maragliano2006}, a brief summary of the steps needed to obtain
Eq.~\eqref{eq:MFEP} is given below to provide the uninitiated reader
with the basic conceptual framework of the string method in collective
variables:
\begin{itemize}
\item[i)] A system of stochastic differential equations in the phase
	space of the original microscopic system (the $6 n_f$-dimensional
	space  given by the vector of coordinates $\bm r$ and velocities
	--momenta expressed in mass reduced coordinates-- ${\bm v} = {\dot
	{\bm r}}$) is devised whose statistically steady solution follow the
	equilibrium probability distribution $m({\bm r}, {\bm v})$ specified
	by the given ensemble \cite{maragliano2006}. Its configurational part
	is the distribution  $m({\bm r}) = \int \mathrm d {\bm v} \; m({\bm r}, {\bm
	v})$ in Eq..~\eqref{landau} and \eqref{metric}.
\item[ii)] The backward Kolmogorov equation (BKE) $L q=0$ associated with the system of stochastic differential equations is identified. BKE is a
partial differential equation for the scalar quantity $q$ in the $6 n_f$-dimensional phase space of the system. $L$ is a linear partial differential operator involving derivatives with respect to both $\bm r$ and $\bm v$, $L = L_{{\bm r}, {\bm v}}$.
When BKE is solved with boundary conditions $q\left({\bm r}, {\bm v}  \right) = 0$ for $\left({\bm r},  {\bm v}\right) \in A$ and
$q\left({\bm r}, {\bm v}  \right) = 1$ for $\left({\bm r}, {\bm v}\right) \in B$, where $A$ and $B$ are the two sets in phase space corresponding to the two metastable macroscopic  states (Cassie and Wenzel, respectively) to be addressed, its solution provides the committor $q\left({\bm r}, {\bm v} \right)$. $q$ gives the probability that the  state $\left({\bm r}, {\bm v}  \right)$ will reach $B$ (the products, in the chemical nomenclature) before reaching $A$ (the reactants). 
The solution of the BKE satisfies a minimum principle for the functional
$
I = \int m\left({\bm r}, {\bm v}\right) \vert L q\vert^2 d {\bm r} d {\bm v} 
$
under the given boundary conditions. The committor is the actual reaction coordinate of the 
transition from $A$ to $B$, and its level surfaces $q\left({\bm r}, {\bm v} \right) = q_0$ are the loci 
of the micro-states with the same progress of the reaction.

\item[iii)] It is assumed that the selected collective variable, namely the coarse-grained density 
field $\theta({\bm r}, {\bm x})$, provides a good description of the transition. 
As a consequence the transition is described as well by minimizing the restriction 
$\tilde I$ of $I$  to the space of functions 
$f(\rho_1, \ldots, \rho_M)$ defined on the coarse grained density field.
This is tantamount to assuming 
$q \simeq f\left[\theta({\bm r}, {\bm x})\right]$.
Two aspects are worth being noted: a) the minimum of $\tilde I$ is now
searched in the space of functions of $M$ variables, $f(\rho_1, \ldots,
\rho_M)$, rather than in that of functions $q$ defined on the $6 n_f$-dimensional
phase space. 
b) The CVs are taken to {\em depend} only on the configuration of the particles and not on their 
velocities ${\bm v}$.
As we shall immediately see, upon minimizing $\tilde I$ with respect to $\theta$, certain averages 
naturally emerge. The functional
\begin{eqnarray}
\nonumber
\tilde I = \int  \mathrm d{\bm r} \mathrm d {\bm v} \; m\left({\bm r}, {\bm v}\right) L f\left[\theta\left( {\bm r }, {\bm x} \right) \right]  L \, f\left[\theta\left({\bm r },{\bm x} \right) \right]=
\\ \nonumber
\int  d {\bm v } \; m_{\bm v }\left( {\bm v}\right) \int \mathrm d {\bm r } \; m_{\bm r }\left( {\bm r}\right) L _{{\bm r}} f \left[\theta\left( {\bm r }, {\bm x} \right) \right] 
L _{{\bm r}} f \left[\theta\left({\bm r },{\bm x} \right) \right]  \, ,
\end{eqnarray}
where the pdf of the relevant ensemble is typically factorized, $m\left({\bm r}, {\bm v} \right) = m_{\bm v}\left( {\bm v} \right)  m_{\bm r}\left( {\bm r} \right)$,
and $L_{\bm r}$ is the part of  $L_{{\bm r}, {\bm v}}$, only involves derivatives with respect to the configuration $\bm r$.
To proceed further,  the explicit expression of $L_{\bm r}$ is needed. For the system we are considering it turns out to be $L_{\bm r} = {\bm v}_i \cdot \partial /\partial {{\bm r}_i}$ (sum is implied on repeated indices). 
\newpage
It follows
\begin{align}
\nonumber
\tilde I = & \int \mathrm  d {\bm v }\;  m_{\bm v }\left( {\bm v}\right) {\bm v}_p \otimes {\bm v}_l :
\int\mathrm  d {\bm r }\;  m_{\bm r }\left( {\bm r}\right) 
\frac{\partial f}{\partial \theta_i} \frac{\partial \theta_i}{\partial {\bm r}_p}
\otimes \frac{\partial \theta_j}{\partial {\bm r}_l} \frac{\partial f}{\partial \theta_j} = 
 \\ \nonumber
 & C \int\mathrm  d {\bm r }\;  m_{\bm r }\left( {\bm r}\right) 
\frac{\partial f}{\partial \theta_i} \frac{\partial \theta_i}{\partial {\bm r}_p}
\cdot \frac{\partial \theta_j}{\partial {\bm r}_p} \frac{\partial f}{\partial \theta_j} =
\\ \nonumber
& C \int\mathrm  d {\bm r } \; m_{\bm r }\left( {\bm r}\right) 
\frac{\partial \theta_i}{\partial {\bm r}_p}
\cdot \frac{\partial \theta_j}{\partial {\bm r}_p} 
\int \mathrm d \rho_1  \ldots \mathrm d \rho_M \;
\frac{\partial f}{\partial \rho_i} 
\frac{\partial f}{\partial \rho_j} 
\\ \nonumber
&\quad\qquad  \qquad  \qquad \qquad  
\delta\left[\rho_1 - \theta_1({\bm r}) \right] \ldots \delta\left[\rho_M - \theta_M({\bm r}) \right] = 
\\ \nonumber
& C \int \mathrm d  \rho_1 \ldots \mathrm d \rho_M \; \mathrm e^{-\beta \Omega(\rho_1, \ldots, \rho_M)}
\frac{\partial f}{\partial \rho_i} 
\frac{\partial f}{\partial \rho_j} 
\\ \nonumber
& \quad \quad
\frac{\displaystyle 
\int  \mathrm d {\bm r } \; m_{\bm r }\left( {\bm r}\right)  \frac{\partial \theta_i}{\partial {\bm r}_p}
\cdot \frac{\partial \theta_j}{\partial {\bm r}_p} \delta\left[\rho_1 - \theta_1({\bm r}) \right] \ldots \delta\left[\rho_M - \theta_M({\bm r}) \right] }{
\displaystyle 
\int \mathrm  d {\bm r } \; m_{\bm r }\left( {\bm r}\right)  \delta\left[\rho_1 - \theta_1({\bm r}) \right] \ldots \delta\left[\rho_M - \theta_M({\bm r}) \right] }
=
\\ \nonumber
& C \int \mathrm d \rho_1 \ldots \rho_m  \; \mathrm e^{-\beta \Omega(\rho_1, \ldots, \rho_M )} \frac{\partial f}{\partial \rho_i} 
\frac{\partial f}{\partial \rho_j}  g_{ij}\left(\rho_1, \ldots, \rho_M \right) \, ,
\end{align}
where $\Omega\left(\rho_1, \ldots, \rho_M \right)$ and
$g_{ij}\left(\rho_1, \ldots, \rho_M \right)$ have been defined in
Eq..~\eqref{landau} and \eqref{metric}, respectively. Note that $\int
\mathrm d \bm v\; m_{\bm v}(\bm v) {\bm v}_p \otimes {\bm v}_l = C \delta_{pl}$ since the velocity of particle $p$ is independent from particle $l$.
\item[iv)] The Euler-Lagrange equations for the restricted functional $\tilde I$ are easily derived as
\begin{eqnarray}
\nonumber
\frac{\partial }{\partial \rho_i} \left(
\mathrm e^{- \beta \Omega\left(\rho_1, \ldots, \rho_M \right)} g_{ij} \frac{\partial f}{\partial \rho_j} \right)= 0 \, ,
\end{eqnarray}
and can be interpreted as the backward Kolmogorov equation of the stochastic differential equation
\begin{eqnarray}
\nonumber
\frac{\mathrm d \rho_i}{\mathrm d \tau} = - g_{ij} \frac{\partial \Omega}{\partial \rho_j} - \frac{1}{\beta} \frac{\partial g_{ij}}{\partial \rho_j} + \sqrt{\frac{2}{\beta}} 
g_{ij}^{1/2} \xi_j \, ,
\end{eqnarray}
where $\xi_s$ is a white noise, $\langle \xi_s\rangle = 0$, $\langle \xi_s(\tau) \xi_r(\tau') \rangle = \delta_{sr} \delta\left(\tau - \tau' \right)$.
\item[v)] When the thermal energy $1/\beta = k_B T$ is small with respect to the typical free 
energy variation along the path (of the order of the
free energy barrier the system has to overcome to undergo the transition which, for rare events, is 
always much larger than $k_B T$)
the leading term in $k_B T$ is negligible and one is left with a stochastic dynamics 
in which the probability of a discrete path connecting the initial and final state is a product of 
Gaussian terms. One can, thus, search for the most probable among these paths, i.e., the path that the transition almost certainly follows:
\begin{eqnarray}
\nonumber
\frac{\mathrm d \rho_i}{\mathrm d \tau} = - g_{ij} \frac{\partial \Omega}{\partial \rho_j} \,  ,
\end{eqnarray}
with boundary conditions $\rho_a \in a$ and $\rho_b \in b$, being $a$ and $b$ the two sets in 
the space of coarse grained density distributions which corresponds to the Cassie and the Wenzel 
state, respectively.  
The problem, as stated in the main text, cannot be easily solved by direct
integration. Reparametrization of the curve $\rho({\bm x})$ in terms of
the arc-length $\lambda$  yields Eq.~\eqref{eq:MFEP} and  the problem can be
reinterpreted as the relaxation of a string
$\rho\left(\bm x; \lambda \right)$ joining the two metastable states
$\rho({\bm x}, 0) = \rho_{a} \in a$ and  $\rho({\bm x},1) = \rho_{b} \in
b$ towards the most probable path. 
\end{itemize}

%%%%%%%%%%%%%%%%%%%%%%%%%%%%%%%%%%%%%%%%%%%%%%%%%%%%%%%%%%%%%
\begin{acknowledgments}
The authors acknowledge fruitful discussion with Francesco Salvadore.
The research leading to these results has received funding from
the European Research Council under the European Union's
Seventh Framework Programme (FP7/2007-2013)/ERC Grant agreement n. [339446].  
We acknowledge PRACE for awarding us access to resource
FERMI based in Italy at Casalecchio di Reno.
\end{acknowledgments}

%%%%%%%%%%%%%%%%%%%%%%%%%%%%%%%%%%%%%%%%%%%%%%%%%%%%%%%%%%%%%
%\bibliographystyle{ieeetr}
\bibliography{biblio}
%%%%%%%%%%%%%%%%%%%%%%%%%%%%%%%%%%%%%%%%%%%%%%%%%%%%%%%%%%%%%

\end{document}